\documentclass[structabstract]{aa}  
\usepackage{graphicx}
\usepackage{txfonts}
\usepackage{natbib}
\bibpunct{(}{)}{;}{a}{}{,} 
\begin{document}
%
\newcommand{\tuc}{\mbox{\rm $J$\,=\,1$-$0}}            
\newcommand{\tdu}{\mbox{\rm $J$\,=\,2$-$1}}         
\newcommand{\ttd}{\mbox{\rm $J$\,=\,3$-$2}}         
\newcommand{\doce}{\rm $^{12}$CO}       
\newcommand{\trece}{\rm $^{13}$CO}      
\newcommand{\gsim}{\raisebox{-.4ex}{$\stackrel{>}{\scriptstyle \sim}$}}
\newcommand{\lsim}{\raisebox{-.4ex}{$\stackrel{<}{\scriptstyle \sim}$}}
\newcommand{\psim}{\raisebox{-.4ex}{$\stackrel{\propto}{\scriptstyle \sim}$}}
\newcommand{\kms}{\mbox{km~s$^{-1}$}}
\newcommand{\mjyb}{\mbox{mJy~beam$^{-1}$}}
\newcommand{\s}{\mbox{$''$}}
\newcommand{\mloss}{\mbox{$\dot{M}$}}
\newcommand{\my}{\mbox{$M_{\odot}$~yr$^{-1}$}}
\newcommand{\ls}{\mbox{$L_{\odot}$}}
\newcommand{\ms}{\mbox{$M_{\odot}$}}
\newcommand{\mm}{\mbox{$\mu$m}}
\def\arcdeg{\hbox{$^\circ$}}
\newcommand{\secp}{\mbox{\rlap{.}$''$}}
\newcommand{\secs}{\mbox{\rlap{.}$^{\rm s}$}}
\newcommand{\um}{\mbox{$\mu$m}}
\newcommand{\h}{$^{\rm h}$}
\newcommand{\m}{$^{\rm m}$}        
   \title{Two short mass-loss events that unveil the binary heart of
Minkowski's Butterfly Nebula\thanks{Based on observations carried out
with the IRAM Plateau de Bure interferometer and 30m
radio-telescope. IRAM is supported by INSU/CNRS (France), MPG
(Germany) and IGN (Spain).}}


   \author{A.\ Castro-Carrizo\inst{1}
          \and
          R.\ Neri\inst{1}
          \and 
	  V.\ Bujarrabal\inst{2}
          \and 
	  O.\ Chesneau\inst{3}
          \and 
	  P.\ Cox\inst{1}
          \and 
	  R.\ Bachiller\inst{4}
          }

   \offprints{ccarrizo@iram.fr}

   \institute{Institut de Radioastronomie Millim\'etrique, 300 rue de la Piscine,
 38406 Saint Martin d'H\`eres, France\\
              \email{ccarrizo@iram.fr,neri@iram.fr,cox@iram.fr}
         \and
              Observatorio Astron\'omico Nacional, Ap 112, E-28803 
Alcal\'a de Henares, Spain\\
             \email{v.bujarrabal@oan.es}
         \and
             UMR 6525 
 Fizeau, Univ. Nice Sophia Antipolis, CNRS, Obs.\ de la C\^{o}te 
d'Azur, Bvd de l'Observatoire, BP4229 F-06304 NICE Cedex 4 \\
             \email{Olivier.Chesneau@oca.eu}
         \and
             Observatorio Astron\'omico Nacional, Alfonso XII N$^o$3, 
E-28014 Madrid, Spain \\
             \email{r.bachiller@oan.es}
             }

   \date{Received 23 December 2011 / Accepted 31 May 2012}

 
  \abstract
   {Studying the appearance and properties of bipolar winds is
     critical to understand the stellar evolution from the AGB to the
     planetary nebula (PN) phase.  Many uncertainties exist regarding
     the presence and role of binary stellar systems, mainly due to
     the deficit of conclusive observational evidences. }
   { We investigate the extended equatorial distribution around the
     early bipolar planetary nebula M 2$-$9 (``Minkowski's Butterfly
     Nebula'') to gather new information on the mechanism of the axial
     ejections.}
   {Interferometric millimeter observations of molecular emission
     provide the most comprehensive view of the equatorial mass
     distribution and kinematics in early PNe.  Here we present
     subarcsecond angular-resolution observations of the $^{12}$CO
     \tdu\ line and continuum emission with the Plateau de Bure
     interferometer.}
   {The data reveal two ring-shaped and eccentric structures at the
     equatorial basis of the two coaxial optical lobes. The two rings
     were formed during short mass-loss episodes ($\sim$ 40 yr),
     separated by $\sim$ 500 yr. Their positional and dynamical
     imprints provide evidence of the presence of a binary stellar
     system at the center, which yields critical information on its
     orbital characteristics, including a mass estimate for the
     secondary of $\lsim$ 0.2 \ms.  The presence of a stellar system
     with a modest-mass companion at the center of such an elongated
     bipolar PN strongly supports the binary-based models, because
     these are more easily able to explain the frequent axisymmetric
     ejections in PNe.}
{}


\maketitle
%

\section{Introduction}

The appearance and shaping of bipolar winds in a stage when stars
transit from the late asymptotic giant branch (AGB) to the planetary
nebula (PN) phase is one of the most intriguing open questions in
stellar evolution. Studying bipolar winds \citep{balickfrank02} is
critical for unraveling this late stellar transition and ultimately
for understanding the formation of PNe, which often show strongly
axisymmetric shapes \citep[e.g.][]{sahai07}. One mechanism that is
supposed to trigger the generation of bipolar winds is the presence of
a binary stellar system at the nebulae's center
\citep{soker01,frankb04,demarco09}. The direct detection of stellar
companions remains an observational challenge, however. Indeed,
central stars of bipolar post-AGB nebulae are rarely found to be
multiple \citep{hrivnak11}.

M 2$-$9 (also known as ``Minkowski's Butterfly Nebula'' or ``the Twin
Jet Nebula'') is a prototypical young planetary nebula that displays a
very elongated bipolar structure \citep[120$\arcsec\times$12$\arcsec$
in size,][]{schwarz97}. It became very popular from the spectacular
images obtained by the Hubble Space Telescope \citep{nasa,balick99} in
the central 60\arcsec-size region of the nebula, which show two
coaxial thin outflows. In the equator, the waist of the nebula is very
slim \citep{nasa}, and its optical image is diffuse due to a
dense torus of gas and dust.

A sequence of observations of ionized gas emission obtained every 2-5
years \citep{doyle00,corradi11} shows a pattern of changes with mirror
symmetry, with respect to the equatorial plane, that rotates around
the symmetry axis of the nebula. A rotating ionizing/exciting beam
from the star is postulated to explain the observed increase in
brightness in a line of knots along the lobe edges. These observations
indicate that there is a binary stellar system at the center
\citep{schmeja01,smith05}, orbiting with a period $\sim$ 90 years
\citep{doyle00,corradi11,livio01,smith2-05}.

\begin{figure*}[h!t]
   \centering
   \includegraphics[angle=0,width=0.99\textwidth]{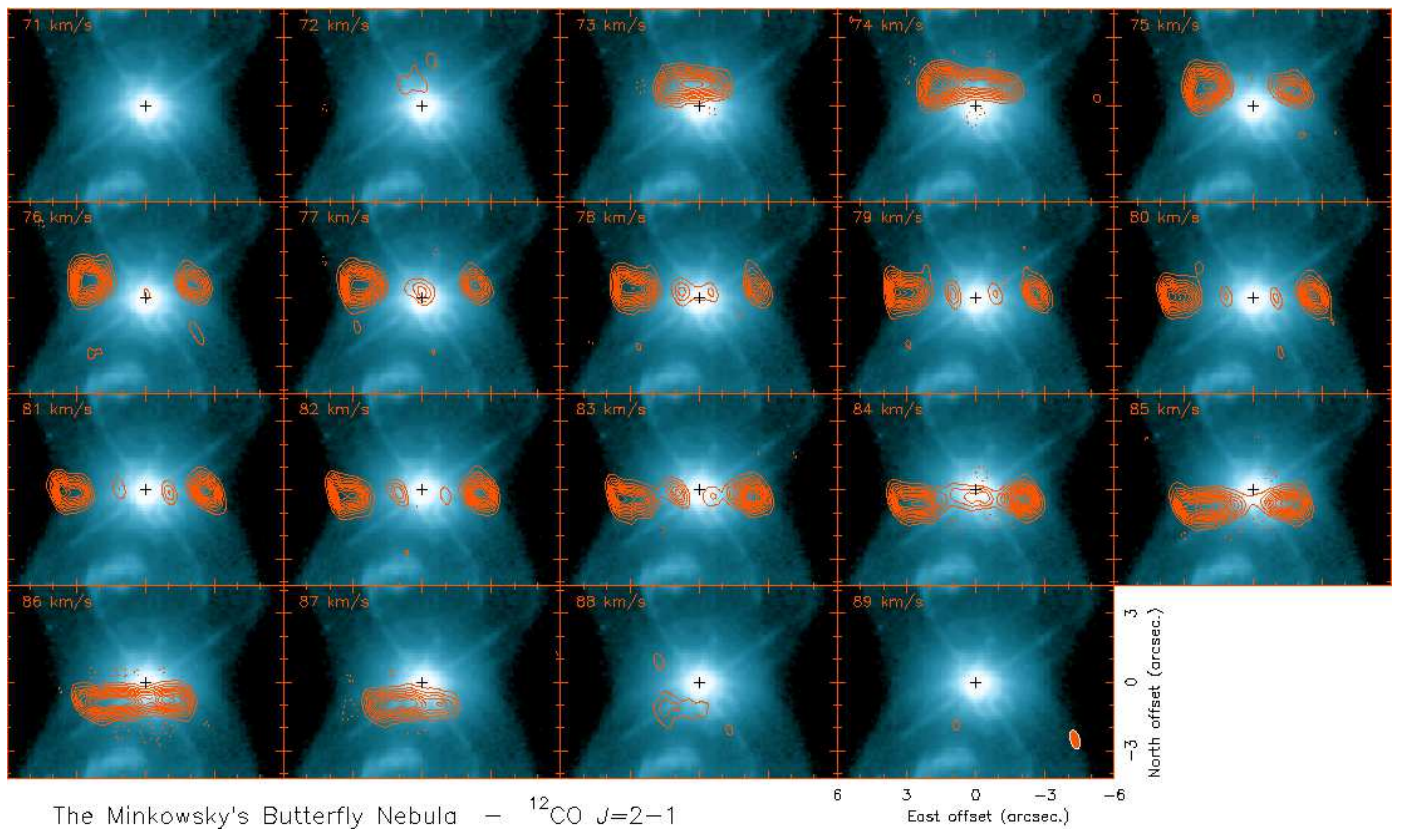}
\caption{Channel maps of the $^{12}$CO \tdu\ line emission toward M
  2$-$9 (in contours) superimposed on the optical image obtained with
  the Hubble Space Telescope \citep[in color scale;][]{nasa}.  The
  center is given by the position of compact continuum emission, here
  subtracted, at the J2000 coordinates RA 17:05:37.958, Dec
  $-$10:08:32.48.  The LSR velocities are specified in the top-left
  corner of each panel. Contours are shown from 4 $\sigma$ with a
  spacing of 5 $\sigma$ (where the root-mean-square noise is 7.9
  \mjyb) in solid line, and at $-$4 $\sigma$ in dashed lines. The
  synthesized beam is 0\farcs86 $\times$ 0\farcs40 at PA = 15\degr\
  and is drawn in the bottom-right corner of the last panel. }
\label{maps}
\end{figure*}

Several theoretical studies have addressed the possible relationship
between nebular bipolarity and binarity of its nucleus, which is one
of the major long-standing problems toward a better understanding of
stellar evolution.  Particularly, for M 2$-$9 it was proposed
\citep{livio01} that a white dwarf star accretes mass from the wind of
an AGB or post-AGB companion, and that a subsequent increase in
magneto-centrifugal forces launches the fast bipolar jets responsible
for the rotating knots. Since the core \citep{lykou11,torres10} is
embedded within a high-density dusty environment, there is no direct
evidence of this scenario yet. On the other hand, the presence of
disks in rotation in the innermost equatorial regions of several young
PNe has been found to be associated to binarity
\citep{bujarrabal05,winckel03}. Several recent works have investigated
such disks using mid-IR high-resolution observations
\citep{matsuura06,chesneau07,lykou11,lagadec11}.

Low-excitation rotational lines of carbon monoxide (CO) are
particularly useful for probing the obscured cores at the center of
evolved stars, including young PNe. They trace the mass distribution,
hindered by extinction in the optical, and provide direct information
on the gas kinematics
\citep{vb01-alas,bujarrabal98,zweigle97,bachiller88}.
\citet{zweigle97} mapped the $^{12}$CO \tdu\ line emission in M 2$-$9
with the Plateau de Bure Interferometer (PdBI). These maps revealed,
with an angular resolution of 3\arcsec $\times$ 5\arcsec, a large
expanding equatorial ring in the nebula waist. \citet{zweigle97}
estimated from $^{12}$CO \tuc\ line emission a mass of $\sim$ 0.01
\ms.
To explore the equatorial mass distribution in M 2$-$9 in more detail,
we carried out new PdBI high angular-resolution observations. We
present in this paper the results of those observations, and analyze
in detail the morphology and kinematics that lead to a new
understanding of the nebula shaping and the central stellar system.

\section{\label{obs} Observations of $^{12}$CO $J=$2$-$1 line and 1\,mm continuum emission}

Observations of $^{12}$CO \tdu\ line emission (at 230.538 GHz) were
performed with the Plateau de Bure interferometer in M\,2$-$9. A track
of observations was obtained in each Aq and Bq array configuration in
February 2009. Channel maps were obtained with a synthetic beam (at
half power) of 0\farcs8 $\times$ 0\farcs4 (P.A.\ 15\degr) in
size. Image synthesis was performed with natural weighting and with
the Hogbom cleaning algorithm.  The results are presented in Fig.\
\ref{maps} with a channel spacing of 1 \kms, though a detailed data
inspection was performed to spectral resolutions of 0.1 \kms. In
Figs.\ \ref{maps} and \ref{integrated} the new M 2$-$9 CO data are
superimposed over the optical image obtained with the Hubble Space
Telescope \citep{nasa}. Astrometry was performed to properly overlay
millimetric and optical data.

Simultaneously, we observed a 1.8GHz bandwidth free of line emission,
which was averaged to detect unresolved 1.3 mm continuum emission of
240 $\pm$ 1 mJy in brightness. Note that this emission was subtracted
in the interferometric visibilities before imaging the CO line
emission presented in this paper. The measured continuum flux is
compatible with a previous analysis of single-dish observations by
\citet{csc98}, where the central compact continuum emission was
deduced to originate from a region of warm dust close to the stellar
system.

In addition to the ring seen in previous studies \citep{zweigle97},
the new observations reveal a second tight inner ring in the waist of
M 2$-$9. The new ring is almost three times smaller and its $^{12}$CO
\tdu\ emission is about four times weaker than that of the outer ring.

Fig.\ \ref{integrated} presents in an inset the velocity-integrated
$^{12}$CO \tdu\ line emission synthesized with a circular beam of
0\farcs4 in size. Only the brightest CO line emission was included to
better separate the closest parts of the two rings (northernmost and
southernmost regions).  As in Fig.\ \ref{maps}, the inset is
superimposed over the HST optical image. The continuum emission is
represented with a red dot.

At first we deduce that the inclination angles of the two equatorial
rings coincide, and are perpendicular to the optical lobes (see
further analysis in Sect.\ \ref{2outflows}). No emission is detected
in the intermediate regions down to a level of 10 \mjyb. The most
extended molecular ring is detached by more than 1\arcsec\ from the
inner one. Its outermost deprojected diameter is 7\farcs2, and the
innermost one 4\farcs2. The easternmost part of the outer ring seems
to reach regions outside the external optical lobes, perhaps with some
flaring, and is thicker than the westernmost part. The inner ring is
2\farcs4 in size, and its easternmost part also presents a moderate
increase in brightness. From the spatial resolution and the
signal-to-noise ratio, we consider our size estimates to be better
than 0\farcs2.

Position-velocity diagrams of the $^{12}$CO \tdu\ line emission in M
2$-$9 along the east-west and north-south directions are shown in
Fig.\ \ref{pv}.


\section{Data analysis}

Two coaxial rings or torus-like structures have been found in
$^{12}$CO \tdu\ line emission laying on the equatorial plane of the
elongated M\,2$-$9 nebula.

\subsection{\label{2outflows} Two outflows in M 2$-$9}

\begin{figure}
   \centering
   \includegraphics[angle=0,width=0.45\textwidth]{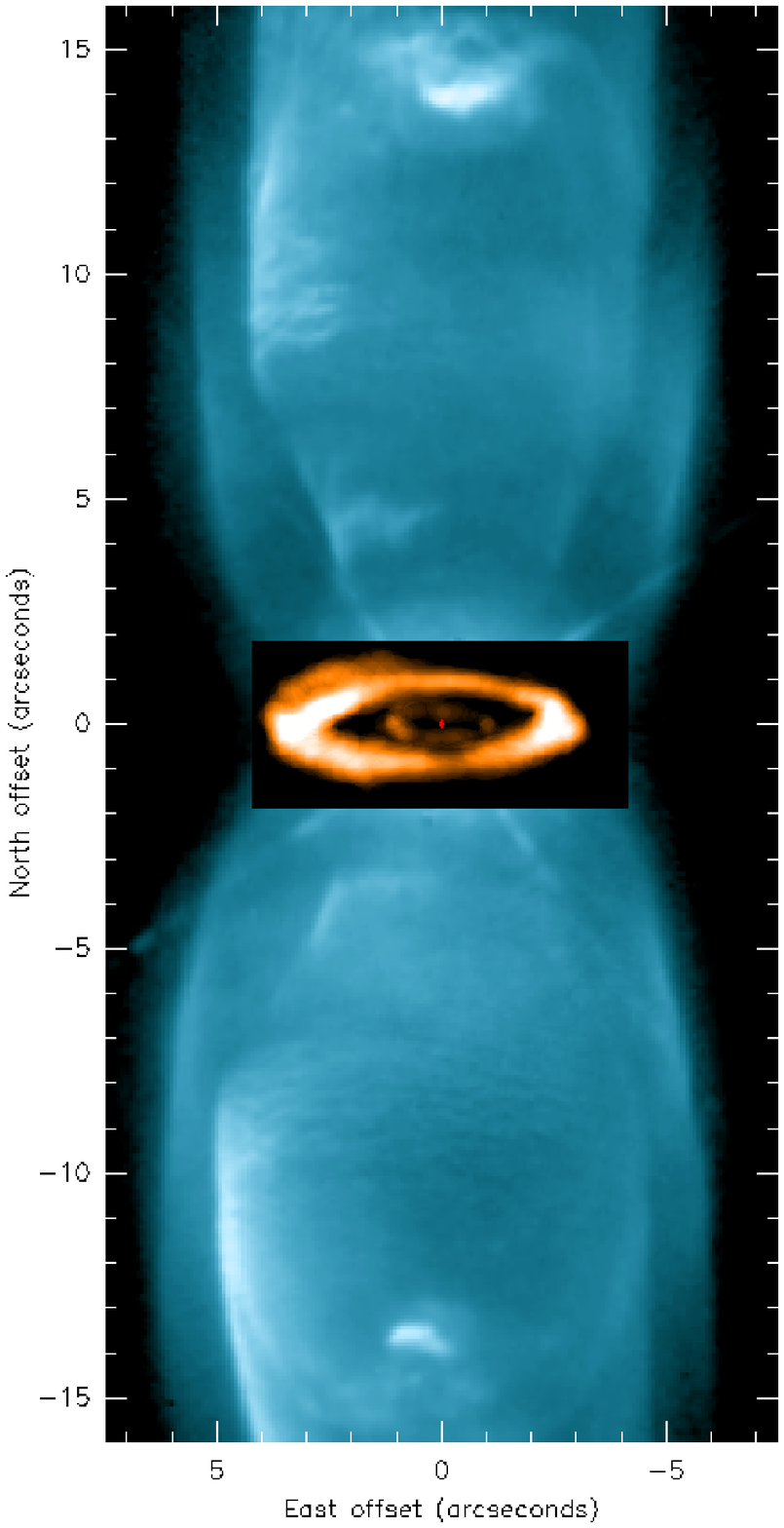}
\caption{Velocity-integrated $^{12}$CO \tdu\ line emission is shown in
the inset, superimposed over the optical image obtained with the
Hubble Space Telescope \citep{nasa}.  Only the brightness distribution
above 60 \mjyb\ (which corresponds to $\sim$ 8 $\sigma$, where
$\sigma$ is the root-mean-square noise) was included to better
separate the close projected image of the two rings. At the center the
continuum emission is represented by a red dot, which traces the
position of the stellar system. More details can be found in Sect.\
\ref{obs}. }
\label{integrated}
\end{figure}

The relation of the two rings and the extended optical lobes is best
shown in Fig.\ \ref{integrated}. This comparison suggests that the two
rings are the equatorial counterparts of the two coaxial optical lobes
(the outer one being more diffuse). The shaping of the rings (or
torus-like structures) and the formation of the bipolar optical
outflows seem hence to result from the same two events of material
ejection.  The low equatorial expansion velocities measured for the
two rings (Sect.\ \ref{kine}) suggest that they likely originate from
an increase in the mass loss when the optical lobes were formed.  Note
that by extrapolating the image of the innermost optical lobes, from
the region where their image becomes diffuse (at $\sim$ 2\arcsec\ from
the equator) to the equatorial plane, we measured that the offset
between the lobes and the rings in the equator must be smaller than
0\farcs4 \citep[the rings being outwards; this relation was
investigated in young PNe by][]{huggins07}.

Assuming that the rings are circular, we estimated the inclination of
their symmetry axes with respect to the plane of the sky from their
projected and actual sizes (along the symmetry and perpendicular
axes). For the two rings we derived a similar inclination of $\sim$
19\degr, with an uncertainty of $\pm$ 2\degr\ due to ring asymmetries.
The northern (southern) part of the ring is approaching (receding
from) us. Therefore, we see the south face of the ring. Similar
inclinations have been derived for the symmetry axis of the bipolar
outflows seen in the optical \citep{solf00,schwarz97}. This confirms
that the rings are accurately placed in the nebula equator.

\subsection{\label{lifetime}The lifetime of the nebula: its formation by 
means of two ejections }

\begin{figure}
   \centering
   \includegraphics[angle=0,width=0.45\textwidth]{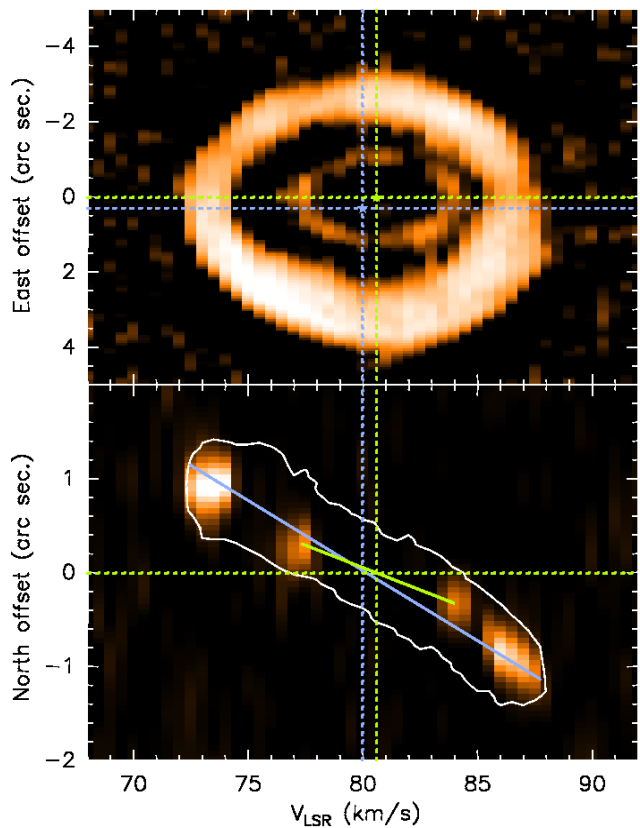}
\caption{Position-velocity diagrams of the $^{12}$CO \tdu\ line
emission in M 2$-$9 along the east-west (on the top) and north-south
(on the bottom) directions.  The green (violet) dotted lines represent
the kinematical and spatial center of the inner (outer) ring.
{\it Top:} Integrated brightness from the whole nebula.
{\it Bottom:} Brightness obtained within a slit along the projected
symmetry axis. With a green (violet) solid line we plot the
position-velocity gradient corresponding to the entire inner (outer)
ring along the north-south direction. A white contour traces the
emission integrated over the complete nebula at low brightness level.
Note that the (V$_{LSR}$) velocity axis shows the expanding gas speed
projected on the line of sight.
A more detailed discussion can be found in Sect.\ \ref{lifetime}.}

\label{pv}
\end{figure}

The position-velocity diagrams presented in Fig.\ \ref{pv} clearly
show that both rings expand. By assuming that their expansion is
isotropic, the systemic velocities corresponding to the large and
small rings are 80.0 and 80.6 \kms.  The systemic velocity associated
to the outer ring hence differs from that of the inner ring by $\sim$
0.6 \kms\ ($\pm$ 0.1 \kms).

In Figs.\ \ref{maps}-\ref{pv} we also see that the centers of both
rings are shifted along the east-west axis.  The inner-ring center
seems to coincide with the position of the stellar system, as traced
by the continuum emission, whereas the center of the larger ring is
offset eastwards by $\sim$ 0\farcs3.
The uncertainty in this estimate mainly comes from departures from
circular symmetry, since the easternmost part of the larger ring is
thicker than the westernmost part. If we only consider the regions
brighter than half the maximum, we estimate that the measured offset
is 0\farcs34 $\pm$ 0\farcs07. Considering the weakest detected
regions, the offset would be 0\farcs31 $\pm$ 0\farcs09.
Note that the uncertainty triggered by the interferometer position
precision is considerably smaller.

These kinematical and spatial differences between the two rings
strongly suggest that the velocity of the mass-losing star was
different when it expelled the two rings.

The ratios between the projected sizes and expansion velocities
presented in the position-velocity gradients along the north-south
direction (in Fig.\ \ref{pv}) provide the times at which the ejections
occurred (multiplied by the tangent of the inclination angle of the
symmetry axis with respect to the sky plane). These different
gradients for the two rings (represented with solid lines in Fig.\
\ref{pv}) would hence be another indication that the rings were formed
at different epochs. For a distance of 650 pc \citep[see ][and
discussion in Sect.\ \ref{distance}]{schwarz97}, we deduce that the
time elapsed between the ejections of the two molecular rings is 500
yr. This value cannot be much lower (by less than 10$\%$) considering
the uncertainties in the inclinations of the rings. The first
ejection, which shaped the outer ring, would have happened about 1400
yr ago. The second ejection, responsible for the inner ring, would
have occurred 900 yr ago. We estimate a maximum uncertainty of 140 yr
in the derived times considering all possible errors.
As discussed there, sizes and times depend proportionally on the
assumed distance; all other results presented in this paper are
distance-independent.

We have already argued that the optical and molecular data show
different parts of the same nebula components, which were formed by
the same (two) events of mass ejection. This implies that the two
lobes seen in the optical are not coeval, but were formed by two
ejection events separated by $\sim$ 500 yr, contrary to what was
concluded in previous studies \citep{smith2-05}.

The measured difference between the systemic velocities of the two
rings is readily explained if the star responsible for the ejections
changed its velocity. This can be justified if the star orbits within
a binary (or multiple) system. The very short time ($\sim$ 500 yr)
elapsed between the two ejections compared to the life of a solar-mass
type star ($\sim$ 10$^9$ yr) strongly suggests that the same late AGB
or post-AGB star was responsible for these events. We do not consider
the remote possibility of having two distinct post-AGB stars to have
formed the two outflows. Moreover, the time lag between the two
outflows is $\sim$ 5 times longer than the binary-orbit period of
$\sim$ 90 yr \citep{doyle00,corradi11,livio01}. This suggests that the
ejections were triggered during a critical phase in the evolution of
one star, such as ultimate and sudden increases in the mass-loss rate
of the late AGB (or post-AGB) star.
The interaction between the stars therefore would not have determined
the time at which the axial ejections happened. This is supported by
the kinematical offset detected between the two rings, which implies
that the ejections occurred at two different relative positions within
the binary orbit (Sect.\ \ref{kine}).

Mass-loss variations have been observed in the envelopes around other
AGB and post-AGB stars with different time scales, from $\sim$ 400 yr
\citep[of $\sim$ 3 10$^{-6}$ \my\ for R Cas; see][]{schoier07,acc10}
to $\sim$ 2000 yr \citep[of $\sim$ 10$^{-4}$ \my\ for CRL\,618 and
IRC\,+10420;][]{csc04,acc07}, some of them being also visible as arcs
or rings in the circumstellar envelopes. In M\,2-9, from the estimate
of ejection duration (Sect.\ \ref{kine0}), we derive a mass-loss rate
of $\sim$ 9 10$^{-5}$ \my\ for the first molecular wind. The mass-loss
rate of the second wind, ejected $\sim$ 500 yr later, is likely lower.
However, in spite of these evidences of mass-loss variations in the
late AGB or early post-AGB phase, the current small sample and
significant differences prevent us from a more detailed analysis.

\subsection{\label{kine} The nebula kinematics}

The large and small rings expand with velocities of 7.8 and 3.9 \kms\
($\pm$ 0.1 \kms). The expansion velocity of 7.8 \kms\ is moderate
compared with that seen in most of the AGB outflows, but 3.9 \kms\ is
a very low velocity compared to AGB shell velocities. Recent studies
of the inner equatorial regions of some post-AGB objects, such as
CRL\,618 \citep{csc04,bujarrabal10} and M\,1-92 \citep{ja07}, show
that equatorial outflows in the very late AGB (or early post-AGB)
phase can expand at these very low velocities, which is also supported
by our data.

There is no sign of deceleration within the rings that could indicate
some interaction with a slower previously expelled halo or surrounding
interstellar medium:
no deceleration position-velocity gradient is detected in the rings,
the outermost layers of both rings present a good circular symmetry,
and the mean expansion velocity is uniform in each ring. In addition,
no CO emission was detected close to the rings or to the rest of
nebula, which would indicate the presence of nearby interstellar
{molecular gas}. The good axis-symmetry of the optical nebula also
supports this interpretation. No circumstellar emission is detected
either between the two rings and, moreover, the innermost ring is
expanding much slower than the previous one, which renders any
interaction between them impossible.
We interpret the data to show that the rings expand freely, and
therefore conclude that times derived from the measured velocities
indeed correspond to times elapsed since the mass ejections.

From the estimated binary period and assuming a stellar mass of the
AGB star $\lsim$ 1 \ms, we deduce a binary distance $\lsim$ 20 AU
(corresponding to $\sim$ 0\farcs03 at a distance of 650 pc), which is
indiscernible in the present data.  The measured position offset
between the ring centers of $\sim$ 0\farcs3 is therefore too large by
a factor $>$ 10 with respect to the orbit size. Therefore the relative
motion of the rings, deduced from the difference in the velocity
centroids (Sect.\ \ref{lifetime}), likely also triggers this offset.
The millimeter continuum emission should correspond to the position of
the binary system (Sect.\ \ref{obs}), which seems to coincide with the
center of the inner ring.

\begin{figure}
   \centering
   \includegraphics[angle=0,width=0.35\textwidth]{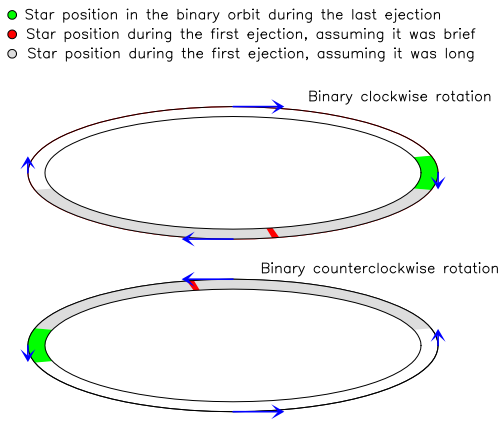}
\caption{Sketch representing the position at which the ejections
occurred in the binary orbit, according to the discussion in Sect.\
\ref{kine} and the two scenarios presented in Sect.\ \ref{kine0}
regarding the first ejection. Note that at first we cannot discern the
sense of the binary rotation, clock- or counterclock-wise (as seen
from the Earth, from the south), and both possibilities are hence
represented in this figure. A clockwise rotation is suggested by
\citet{corradi11} and also by the position-velocity gradient detected
in the southernmost part of the outer ring (see Sect.\ \ref{kine2}). }
\label{binary}
\end{figure}

In Fig.\ \ref{pv} we see that the center of the inner ring remains
very close to the binary position (from our perspective), a possible
displacement in the last 900 yr being smaller than our uncertainty in
the determination of the ring center, which is estimated to be $\sim$
0.06\arcsec\ due to small ring inhomogeneities \citep[slightly larger
than the position uncertainty, $<$ 0.04\arcsec,][]{reid88}.
Accordingly we deduce that the inner ring should be moving in a
direction closer to the line of sight than 12\degr\ (independently of
the distance, by adopting a binary rotation speed of 1 \kms). 
The last ejection happened therefore when the star was moving in a
direction very close to the light of sight, in the easternmost or the
westernmost part of the orbit.

The relative offset between the two rings indicates that when the
first ejection happened, the star was moving eastwards. Their
systemic-velocity offset suggests that the last ejection occurred when
the star was receding. However, at this point we cannot distinguish
the sense of the binary rotation. If clockwise (as seen from the
Earth, from the south), the first ejection would have happened in the
semi-orbit that is farther from us and the last one in the westernmost
part of the binary orbit. If counterclockwise, the first ejection
would have occurred in the section of the orbit that is closer to us,
and the last one in the easternmost part. These two scenarios are
represented in Fig.\ \ref{binary}. \citet{corradi11} deduced that the
observed east-to-west motion of the bright nebular knots takes place
in the side of the bulbs facing the Earth. This would indicate that
the binary presents a clockwise orbit (upper plot in Fig.\
\ref{binary}). Accordingly, we can determine the position of the stars
when the ejections took place, as shown in Sect.\ \ref{kine2}. There,
additional discussion supports the clockwise orbital scenario.

\section{\label{distance}Distance}

\begin{figure}
   \centering
   \includegraphics[angle=0,width=0.53\textwidth]{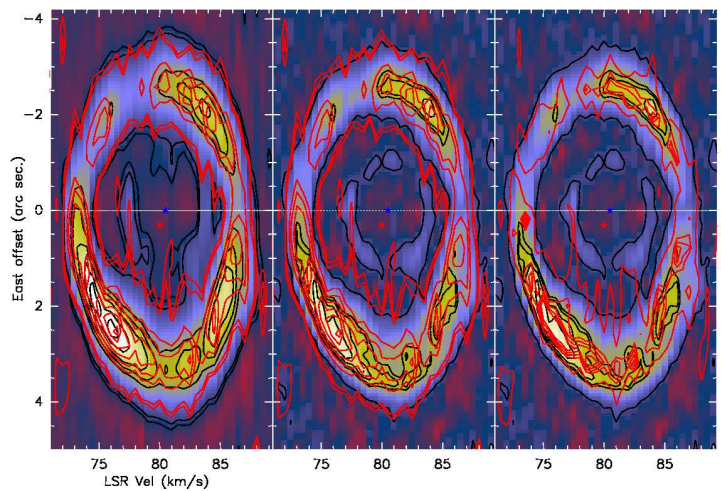}
\caption{
East-west position-velocity diagrams (similar to those in Fig.\
\ref{pv}) to compare the observations performed in 1997 (in red
contours) with the new data observed in 2009 (in color scale and black
contours). From left to right data are synthesized with different
beams.
{\it Left:} All contours and the image are synthesized with the beam
corresponding to the observations obtained in 1997 (of
2$\farcs$38$\times$1$\farcs$04 in size, PA 7\degr).
{\it Center:} Contours are plotted with their respective original
beams, of 1$\farcs$04 and 0$\farcs$4 in size in the east-west
direction for the 1997 and 2009 observations, respectively.
{\it Right:} All contours and the image are synthesized with the beam
obtained in 2009.
A detailed analysis is performed in Sect.\ \ref{distance}. }
\label{dist}
\end{figure}

We adopted a distance of 650 pc, a value so far mostly considered in
the literature for M\,2$-$9 that reasonably agrees with the tentative
ring expansion presented in this section. Recently \citet{corradi11}
proposed a farther distance of 1,300 pc.

To investigate the ring-expansion proper motions, a comparison was
performed between observations obtained in 1997 (unpublished data by
Neri et al.) and the new data presented in this paper. In Fig.\
\ref{dist} we compare the east-west position-velocity diagrams
(equivalent to that in Fig.\ \ref{pv}) for the two data sets,
synthesized with different beams. A correlation analysis of the
brightest parts in the outer ring clearly suggests expansion proper
motion over the elapsed time. From the compared images in Fig.\
\ref{dist}, we estimate a diameter increase that roughly ranges from
0.10\arcsec\ to 0.24\arcsec.
We have performed a similar analysis by fitting the data in the
$uv$-plane.  The two data sets were fitted at the central channel by
two Gaussian functions of FWHP 1\farcs17 and 0\farcs81, for the
eastern and western ring components, their positions being free
fitting parameters.  In this way we estimate that the ring diameter
increased by 0\farcs25 $\pm$ 0\farcs11.

From the measured expansion velocity (7.8 \kms) and the time elapsed
between the two observations (11.5 yr), we calculated the distance to
M 2$-$9 with the formula $d$(pc) =
36.9/$diameter\_increase$(\arcsec). Accordingly, from the
proper-motion estimates we deduce that the distance to M\,2$-$9 ranges
from 100 to 300 pc. By arbitrarily increasing our uncertainty to
0\farcs20, the distance would range from 80 to 800 pc. Therefore,
though we cannot perform a precise distance estimate, it seems
improbable that the distance is larger than 1,000 pc.

Because our detection of proper motion is only tentative, we have
analyzed the uncertainty in the value derived by \citet{corradi11}.
Distances shorter than 1,000 pc should not be discarded either from
their analysis of the proper motions in reflected light, particularly
considering the observational uncertainties and the inconsistencies in
the measured offsets for the two lobes.

The choice of distance has an impact on the main known nebula
characteristics.  The estimated sizes and times in particular are
proportional to the distance. For a distance twice larger \citep[as
proposed by ][]{corradi11}, the total nebula extent \citep[as mapped
by ][]{schwarz97} would be $\sim$ 2 10$^{18}$ cm, about three times
larger than usual PNe sizes \citep[e.g.][]{bujarrabal88}. M 2$-$9
would be one of the largest known PNe. Also, the two ejections would
be separated by $\sim$ 1,000 yr, the first one would have occurred
2,800 yr ago. We note that these times are extraordinary long from
what we expect in the AGB-to-PNe evolution, which is estimated to last
$\sim$ 500 - 2,000 yr \citep{bujarrabal88}. This is particularly
unexpected if we consider that M 2$-$9 is not thought to be a
developed PN but an early PN \citep{acc01-iso}, based on the estimate
of its effective stellar temperature \citep{calvet78,swings79}.

In view of the presented data comparison, the above discussion, and
the uncertainties we have, we prefer to adopt a distance of 650
pc. This value is so far mostly adopted in the literature, and
facilitates a comparison with previous results.

However, we also note that all the main results obtained in this paper
are independent of the adopted distance. Regarding the velocities,
they are either directly measured or derived by the ratio of sizes and
times. Therefore estimated speeds, and consequently masses, are
distance-independent (Sect.\ \ref{kine0}). The inclination angle of
the nebula symmetry axis is also derived from angular sizes (Sect.\
\ref{2outflows}). For the estimate of the direction along which the
inner ring moves (Sect.\ \ref{kine}) the distance dependence cancels
out with the ejection lifetime. The angular size of the binary orbit
would be smaller by adopting 1,300 pc, $\sim$ 0.01\arcsec, which would
make it more difficult to resolve the binary with the current data
(Sect.\ \ref{kine}), again supporting our interpretation.

\section{\label{kine0}Kinematics of the binary orbit}

From the properties of the two rings we can estimate the modulus of
the orbital velocity of the star that ejected the rings, $V$. For this
we assume that the orbit is circular, that $V$ remained constant
during the whole mass-loss phase, and that the same star ejected both
rings.  We note that no time variations have been reported in the
rotation pattern seen in optical observations, and associated to the
orbital rotation itself. Hence, no high eccentricities are expected in
the binary orbit \citep[as concluded by ][]{doyle00,corradi11}, which
supports our main assumptions.
Here we derive properties of the binary rotation from parameters
directly measured from the observations, and the mentioned
assumptions. 
To start, note that the speed difference between the two rings allows
us to give a lower limit for the orbital velocity of 0.31 \kms\
(obtained by deprojecting the measured velocity difference, divided by
two).

Below, we refer to axis $x$ as the axis perpendicular to the east-west
direction within the equatorial plane, passing through the continuum
position, its projection in the sky plane being parallel to that of
the nebular axis. Axis $y$ is perpendicular to axis $x$ within the
equatorial plane, parallel to the east-west direction.

\subsection{\label{kine1} Very short mass-loss event assumption}

In a first approach, we assume that the rings were ejected during
phases much shorter than the orbital period. From the observations, we
deduce 1) A position offset between the ring centers of 0\farcs34
($\pm$ 0\farcs07) in axis $y$, which results from the relative motion
of the rings. 2) The second inner ring is moving in a direction closer
than 12\degr\ to that of axis $x$, which is an upper limit. 3) The
relative velocity between the two rings has a projection in the line
of sight equal to 0.60 \kms\ ($\pm$ 0.10 \kms). From point (2) we can
assume that the last ejection took place when the star was moving in
the direction of axis $x$, so that the velocity of the inner ring was
such that $V^i_y$ $\sim$ 0 and $V^i_x$ $\sim$ $V$ [Eq.\ E1] (where $V$
is the modulus of the binary rotation velocity). Then, by considering
item (1) the offset of 0\farcs34 in axis $y$ is only due to the
velocity of the outer ring, i.e.  $V^o_y$ = 0\farcs34/$T_o$ = 0.74
\kms\ ($\pm$ 0.23 \kms) [Eq.\ E2], where $T_o$ is the time from the
first ejection ($\sim$ 0.45\arcsec/\kms\ from the gradient shown in
Fig.\ \ref{pv}, which multiplied by the distance gives $\sim$ 1400
yr). We hence obtain ($V^o_x$)$^2$ = ($V$)$^2$ - (0.74 \kms)$^2$ [Eq.\
E3].  Finally, from point (3) we deduce that $V^o_x$ - $V^i_x$ = 0.63
\kms\ (where 0.63 \kms\ is estimated by deprojection toward the
equatorial plane of the measured 0.60 \kms). By using Eq.\ E1 we
obtain $V^o_x$ = $V$ - 0.63 \kms\ [Eq.\ E4]. From the last two
equations (E3 and E4) we deduce that $V$ = 0.75 \kms\ ($\pm$ 0.50
\kms). From $V^o_x$ and $V$ we also deduce that the ejection of the
outer ring happened from an orbital point at 17\degr\ from the
intersection with axis $x$.
The velocity-vector direction was hence close to that of axis $y$,
though this angle estimate is uncertain within the southern
semi-orbit.
We finally note that assuming ejections shorter than they actually are
would lead to underestimated velocities.

We can also derive the systemic velocity of the binary stellar system
in M 2$-$9 by correcting the measured systemic velocities of the rings
by the derived stellar velocity at the time of the ejections. This is
particularly straightforward from the inner-ring velocity. We obtain a
stellar speed of 79.8 \kms\ ($\pm$ 0.5 \kms), which is slower and
hence agrees better with the values recently suggested from optical
observations \citep{torres10}. The fact that this value is lower than
the outer-ring systemic velocity would indicate that the first
ejection would have happened when the star was receding with respect
to us within the binary orbit, as represented in Fig.\ \ref{binary},
though moving close to the east-west direction.

\subsection{\label{kine2} Longer mass-loss event assumption}

\begin{figure}[b]
   \centering
   \includegraphics[angle=0,width=0.25\textwidth]{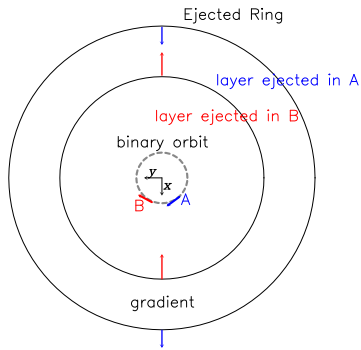}
\caption{Sketch to visualize the origin of the velocity gradient
expected in one side of the ring, if its shaping took place for a
fraction of the binary orbit period. According to the scenario
described in Sect.\ \ref{kine2}, in the sourthern part of the ring the
outermost layers would expand faster than the innermost ones. }
\label{gradient}
\end{figure}

Let us suppose that the first ejection happened over a longer time,
during a significant fraction of the orbital period.
From what is derived in Sect.\ \ref{kine1}, we can suppose that the
oldest ring was ejected when the star was moving basically in the
direction parallel to the east-west axis. The ejection happened when
the star was describing an angle 2$\alpha$ in its orbit. The average
velocity of the outer ejected ring in the $y$ direction is then

$\langle V^o_y \rangle = \frac{\int_{-\alpha}^{\alpha} V cos\sigma \mathrm{d}\sigma}{2 \alpha} = 
\frac{V sin\alpha}{\alpha} $ \,\,\,\,\,\, [Eq.\ E5].

\noindent
In previous estimates (Eq.\ E2) we derived that $\langle V^o_y \rangle
\sim$ 0.7 \kms\ ($\pm$ 0.2 \kms). Note that because of the longer
mass-loss, we would expect an acceleration gradient in the ring along
the perpendicular axis ($x$) whose extreme values would be separated
by $\Delta V_x = 2 V sin\alpha$ [Eq.\ E6]. This would be because the
gas expelled during the first part of the ejection phase would move
faster in the expansion direction ($x$ in our case) than the gas
ejected subsequently, as a consequence of the change in the
$\overrightarrow{V}$ direction. This gradient should only be observed
in the ring part corresponding to the same angle interval described by
the star in its orbit during the ejection. Also, a compression effect
should have happened at the opposite side of the ring (at 180\degr)
due to the contrary relative movement between the different ring
layers (see a descriptive sketch in Fig.\ \ref{gradient}).

We did indeed detect a small acceleration position-velocity
gradient along the axis $x$ in the southernmost part of the
north-south diagram, shown in Fig.\ \ref{pv}: the outermost layers of
the outer ring expand faster than the innermost layers.
Indeed, similar acceleration gradients are detected in the north-south
position-velocity diagrams obtained in the central east-west
3\arcsec-size section.  Considering the uncertainties we could have in
our estimates of the gradient (regarding the beam size and the
comparison with the northern part) and also projection effects, we
estimate that $\Delta V_x =$ 2 \kms\ ($\pm$ 1 \kms). As predicted, we
detected no sign of a velocity gradient in the northernmost side of
the outer ring.  From Eqs.  E5 and E6 we derive $\alpha \sim $ 80\degr
($\pm$ 60\degr), the ejection takes hence place during about half of
an orbital period ($\pm \alpha$) and $V$ would be $\sim$ 1\kms\ $\pm$
1\kms, and therefore ranging from 0.3 (hard limit from the measured
velocities; Sect.\ \ref{kine0}) to 2 \kms.

If our interpretation on the observed relative gradient in the
southern part of the outermost ring is correct, this would be an
additional indication that the ejection took place when the star was
somewhere in the southernmost binary semi-orbit, suggesting a binary
clockwise rotation (as shown in Fig.\ \ref{binary}). As described in
Sect.\ \ref{kine}, this agrees with what was already suggested by
\citet{corradi11}.

We estimate this second scenario, based on the gradient detected in
the ring southern part, to be more realistic, and hence to allow a
better characterization of the first ejection and of the orbital
kinematics.  We note, however, that a deeper analysis of this gradient
is limited by the angular resolution in the north-south direction.

From the deduced orbital velocity we estimate a systemic velocity for
the binary stellar system of $\sim$ 79.6 \kms, which also remains
compatible with \citet[][see previous discussion in Sect.\
\ref{kine1}]{torres10}.
In addition, from the duration estimates we derive that the first
ejection took place during $\sim$ 40 yr ($\pm$ 30 yr).  For the inner
ring, the determination of the ejection time is more difficult, but
must anyhow be considerably shorter considering its narrowness.

\subsection{Implications for the companion mass}

We can estimate the mass of the companion from the derived orbital
velocity. For this we assume again a circular binary orbit and an
orbital period, $P$, of $\sim$ 90 yr
\citep{corradi11,livio01,doyle00}. From the gravitational equations
associated to the movement of two bodies we obtain that the modulus of
the orbital velocity of the primary star (1: presumably the most
massive and responsible for the mass ejection) is in solar units:
$V_1 (\kms) = 30 \times \left(\frac{M_{tot}}{P}\right)^{1/3}
\frac{M_2}{M_{tot}}$ [Eq.\ E7], where $M_{tot}$ is the total mass and
$M_2$ that of the secondary.

By introducing the derived $V$ in Eq.\ E7, we find that the mass of
the secondary must be very small. Indeed, for a mass of the primary
(late AGB) star of $\sim$ 1 \ms\ the mass of the companion is $M_2$
$\sim$ 0.1 $-$ 0.2 \ms. If we adopt a different $M_{tot}$ by a factor
2, the mass of the companion would change by $\pm$ 0.1
\ms. Considering that the total estimated circumstellar mass is
$\lsim$ 0.1 \ms\ \citep[including ionized and atomic gas,
][]{acc01-iso}, a mass for the primary much larger than 2 \ms\ does
not seem very likely. However, note that if $M_{tot}$ would be $\sim$
4 \ms, the mass of the companion would be $\sim$ 0.4 \ms.  In all the
cases the companion would be considerably smaller than the primary
star.

\section{Conclusions}

We presented subarcsecond-resolution mapping of $^{12}$CO \tdu\ line
emission in the proto-planetary nebula M\,2$-$9. Our data show two
independent expanding rings (or torus-like structures) in the nebula
equator. Continuum emission has been detected, likely tracing the
position of the stellar system.

The centers of the two rings are offset by 0.34\arcsec, and their
systemic velocities differ by 0.6 \kms. This is solid evidence that
there is a binary system at the center of M\,2$-$9. In this scenario,
the two rings were very probably ejected by one of the stars, in a
very late AGB or very early post-AGB phase. The different velocities
of the star in its orbit when the material shaping the rings was
expelled explain the measured position and velocity offsets between
the rings. At the same time, these two mass-loss events likely gave
rise to the two coaxial lobes seen in infrared and optical images The
first ejection happened 1400 yr ago, the second 500 yr later.
We suggest that, after the material was expelled by the star, the
axial lobes were shaped, accelerated, by interaction with post-AGB
fast and collimated jets. The interaction effects are still visible in
the bright shock-excited knots detected in the visible but did not
affect the dynamics of the equatorial rings, in agreement with current
theoretical ideas of PN shaping (see Sect.\ 1).
Although most of the material in the equatorial rings and the axial
lobes was probably ejected by the primary (in a late AGB or early
post-AGB phase), the structure and movements of the knots indicate
that the collimated exciting jets came from a compact, less bright
secondary \citep{corradi11}.

A distance of 650 pc was adopted in this paper, which agrees better
with the tentative detection of proper motion we presented here. None
of our main results (except for lifetimes) depend on the adopted
distance.

By analyzing the kinematics of the CO-emitting equatorial rings we
estimated the positions at which the two ejections took place in the
binary orbit. With some few simple assumptions, we derived an orbital
velocity of $\sim$ 1 \kms. From the equations of the orbital movement
of two objects, we obtained a very small mass for the secondary,
$\lsim$ 0.2 \ms\ ($\pm$ 0.1 \ms, by assuming that the mass of the
primary star is between 0.5 and 2 \ms). We suggested that the binary
system is composed of a red giant and a dwarf. We estimated that they
would be separated by 20 $\pm$ 5 AU (for the interval of total masses
considered above), which is too large for the stars to have undergone
a common envelope phase, or for being interacting symbiotic systems
\citep{symbio1,symbio2}.

The strongly axisymmetric nebula and very collimated jets around the
binary system convincingly suggest that the multiple nature of the
stellar component played a relevant role in the ejection and shaping
of M\,2$-$9.  The significance of this low-mass companion to shape the
extended and elongated bipolar nebula seen in M 2$-$9 has important
consequences on our understanding of the mechanisms that drive the
shaping of asymmetrical planetary nebulae, since the existence of
companions is increasingly more probable when the requirements for
their masses are relaxed.

\begin{acknowledgements}
      The data here presented were reduced and analyzed entirely by
      using the different packages available in the GILDAS software
      (http://www.iram.fr/IRAMFR/GILDAS). This project has been
      partially supported by the Spanish MICINN, program CONSOLIDER
      INGENIO 2010, grant ``ASTROMOL" (CSD2009-00038).
\end{acknowledgements}

\bibliographystyle{aa}
\bibliography{mnemonic,castrocarrizo}

\begin{thebibliography}{42}
\expandafter\ifx\csname natexlab\endcsname\relax\def\natexlab#1{#1}\fi

\bibitem[{{Alcolea} {et~al.}(2007){Alcolea}, {Neri}, \& {Bujarrabal}}]{ja07}
{Alcolea}, J., {Neri}, R., \& {Bujarrabal}, V. 2007, \aap, 468, L41

\bibitem[{{Bachiller} {et~al.}(1988){Bachiller}, {Gomez-Gonzalez},
  {Bujarrabal}, \& {Martin-Pintado}}]{bachiller88}
{Bachiller}, R., {Gomez-Gonzalez}, J., {Bujarrabal}, V., \& {Martin-Pintado},
  J. 1988, \aap, 196, L5

\bibitem[{{Balick}(1999)}]{balick99}
{Balick}, B. 1999, in Astronomical Society of the Pacific Conference Series,
  Vol. 188, Optical and Infrared Spectroscopy of Circumstellar Matter, ed.
  {E.~Guenther, B.~Stecklum, \& S.~Klose}, 241--+

\bibitem[{{Balick} \& {Frank}(2002)}]{balickfrank02}
{Balick}, B. \& {Frank}, A. 2002, \araa, 40, 439

\bibitem[{{Balick} {et~al.}(1997){Balick}, {Icke}, {Mellema}, \& {NASA}}]{nasa}
{Balick}, B., {Icke}, V., {Mellema}, G., \& {NASA}. 1997, in NASA press
  release, ed. NASA, NASA press release, GRIN DataBase Number:
  GPN--2000--000953

\bibitem[{{Bujarrabal} {et~al.}(1998){Bujarrabal}, {Alcolea}, {Sahai},
  {Zamorano}, \& {Zijlstra}}]{bujarrabal98}
{Bujarrabal}, V., {Alcolea}, J., {Sahai}, R., {Zamorano}, J., \& {Zijlstra},
  A.~A. 1998, \aap, 331, 361

\bibitem[{{Bujarrabal} {et~al.}(2010){Bujarrabal}, {Alcolea}, {Soria-Ruiz},
  {Planesas}, {Teyssier}, {Marston}, {Cernicharo}, {Decin}, {Dominik},
  {Justtanont}, {de Koter}, {Melnick}, {Menten}, {Neufeld}, {Olofsson},
  {Schmidt}, {Sch{\"o}ier}, {Szczerba}, {Waters}, {Quintana-Lacaci},
  {G{\"u}sten}, {Gallego}, {D{\'{\i}}ez-Gonz{\'a}lez}, {Barcia},
  {L{\'o}pez-Fern{\'a}ndez}, {Wildeman}, {Tielens}, \& {Jacobs}}]{bujarrabal10}
{Bujarrabal}, V., {Alcolea}, J., {Soria-Ruiz}, R., {et~al.} 2010, \aap, 521,
  L3+

\bibitem[{{Bujarrabal} {et~al.}(2005){Bujarrabal}, {Castro-Carrizo}, {Alcolea},
  \& {Neri}}]{bujarrabal05}
{Bujarrabal}, V., {Castro-Carrizo}, A., {Alcolea}, J., \& {Neri}, R. 2005,
  \aap, 441, 1031

\bibitem[{{Bujarrabal} {et~al.}(2001){Bujarrabal}, {Castro-Carrizo}, {Alcolea},
  \& {S{\'a}nchez Contreras}}]{vb01-alas}
{Bujarrabal}, V., {Castro-Carrizo}, A., {Alcolea}, J., \& {S{\'a}nchez
  Contreras}, C. 2001, \aap, 377, 868

\bibitem[{{Bujarrabal} {et~al.}(1988){Bujarrabal}, {Gomez-Gonzalez},
  {Bachiller}, \& {Martin-Pintado}}]{bujarrabal88}
{Bujarrabal}, V., {Gomez-Gonzalez}, J., {Bachiller}, R., \& {Martin-Pintado},
  J. 1988, \aap, 204, 242

\bibitem[{{Calvet} \& {Cohen}(1978)}]{calvet78}
{Calvet}, N. \& {Cohen}, M. 1978, \mnras, 182, 687

\bibitem[{{Castro-Carrizo} {et~al.}(2001){Castro-Carrizo}, {Bujarrabal},
  {Fong}, {Meixner}, {Tielens}, {Latter}, \& {Barlow}}]{acc01-iso}
{Castro-Carrizo}, A., {Bujarrabal}, V., {Fong}, D., {et~al.} 2001, \aap, 367,
  674

\bibitem[{{Castro-Carrizo} {et~al.}(2007){Castro-Carrizo}, {Quintana-Lacaci},
  {Bujarrabal}, {Neri}, \& {Alcolea}}]{acc07}
{Castro-Carrizo}, A., {Quintana-Lacaci}, G., {Bujarrabal}, V., {Neri}, R., \&
  {Alcolea}, J. 2007, \aap, 465, 457

\bibitem[{{Castro-Carrizo} {et~al.}(2010){Castro-Carrizo}, {Quintana-Lacaci},
  {Neri}, {Bujarrabal}, {Sch{\"o}ier}, {Winters}, {Olofsson}, {Lindqvist},
  {Alcolea}, {Lucas}, \& {Grewing}}]{acc10}
{Castro-Carrizo}, A., {Quintana-Lacaci}, G., {Neri}, R., {et~al.} 2010, \aap,
  523, A59+

\bibitem[{{Chesneau} {et~al.}(2007){Chesneau}, {Lykou}, {Balick}, {Lagadec},
  {Matsuura}, {Smith}, {Spang}, {Wolf}, \& {Zijlstra}}]{chesneau07}
{Chesneau}, O., {Lykou}, F., {Balick}, B., {et~al.} 2007, \aap, 473, L29

\bibitem[{{Corradi} {et~al.}(2011){Corradi}, {Balick}, \&
  {Santander-Garc{\'{\i}}a}}]{corradi11}
{Corradi}, R.~L.~M., {Balick}, B., \& {Santander-Garc{\'{\i}}a}, M. 2011, \aap,
  529, A43+

\bibitem[{{De Marco}(2009)}]{demarco09}
{De Marco}, O. 2009, \pasp, 121, 316

\bibitem[{{Doyle} {et~al.}(2000){Doyle}, {Balick}, {Corradi}, \&
  {Schwarz}}]{doyle00}
{Doyle}, S., {Balick}, B., {Corradi}, R.~L.~M., \& {Schwarz}, H.~E. 2000, \aj,
  119, 1339

\bibitem[{{Frank} \& {Blackman}(2004)}]{frankb04}
{Frank}, A. \& {Blackman}, E.~G. 2004, \apj, 614, 737

\bibitem[{{Hrivnak} {et~al.}(2011){Hrivnak}, {Lu}, {Bohlender}, {Morris},
  {Woodsworth}, \& {Scarfe}}]{hrivnak11}
{Hrivnak}, B.~J., {Lu}, W., {Bohlender}, D., {et~al.} 2011, \apj, 734, 25

\bibitem[{{Huggins}(2007)}]{huggins07}
{Huggins}, P.~J. 2007, \apj, 663, 342

\bibitem[{{Lagadec} {et~al.}(2011){Lagadec}, {Verhoelst}, {M{\'e}karnia},
  {Su{\'a}eez}, {Zijlstra}, {Bendjoya}, {Szczerba}, {Chesneau}, {van Winckel},
  {Barlow}, {Matsuura}, {Bowey}, {Lorenz-Martins}, \& {Gledhill}}]{lagadec11}
{Lagadec}, E., {Verhoelst}, T., {M{\'e}karnia}, D., {et~al.} 2011, \mnras, 417,
  32

\bibitem[{{Livio} \& {Soker}(2001)}]{livio01}
{Livio}, M. \& {Soker}, N. 2001, \apj, 552, 685

\bibitem[{{Lykou} {et~al.}(2011){Lykou}, {Chesneau}, {Zijlstra},
  {Castro-Carrizo}, {Lagadec}, {Balick}, \& {Smith}}]{lykou11}
{Lykou}, F., {Chesneau}, O., {Zijlstra}, A.~A., {et~al.} 2011, \aap, 527, A105+

\bibitem[{{Matsuura} {et~al.}(2006){Matsuura}, {Chesneau}, {Zijlstra}, {Jaffe},
  {Waters}, {Yates}, {Lagadec}, {Gledhill}, {Etoka}, \&
  {Richards}}]{matsuura06}
{Matsuura}, M., {Chesneau}, O., {Zijlstra}, A.~A., {et~al.} 2006, \apjl, 646,
  L123

\bibitem[{{Miko{\l}ajewska}(2007)}]{symbio1}
{Miko{\l}ajewska}, J. 2007, Baltic Astronomy, 16, 1

\bibitem[{{Podsiadlowski} \& {Mohamed}(2007)}]{symbio2}
{Podsiadlowski}, P. \& {Mohamed}, S. 2007, Baltic Astronomy, 16, 26

\bibitem[{{Reid} {et~al.}(1988){Reid}, {Schneps}, {Moran}, {Gwinn}, {Genzel},
  {Downes}, \& {Roennaeng}}]{reid88}
{Reid}, M.~J., {Schneps}, M.~H., {Moran}, J.~M., {et~al.} 1988, \apj, 330, 809

\bibitem[{{Sahai} {et~al.}(2007){Sahai}, {Morris}, {S{\'a}nchez Contreras}, \&
  {Claussen}}]{sahai07}
{Sahai}, R., {Morris}, M., {S{\'a}nchez Contreras}, C., \& {Claussen}, M. 2007,
  \aj, 134, 2200

\bibitem[{{S{\'a}nchez Contreras} {et~al.}(1998){S{\'a}nchez Contreras},
  {Alcolea}, {Bujarrabal}, \& {Neri}}]{csc98}
{S{\'a}nchez Contreras}, C., {Alcolea}, J., {Bujarrabal}, V., \& {Neri}, R.
  1998, \aap, 337, 233

\bibitem[{{S{\'a}nchez Contreras} {et~al.}(2004){S{\'a}nchez Contreras},
  {Bujarrabal}, {Castro-Carrizo}, {Alcolea}, \& {Sargent}}]{csc04}
{S{\'a}nchez Contreras}, C., {Bujarrabal}, V., {Castro-Carrizo}, A., {Alcolea},
  J., \& {Sargent}, A. 2004, \apj, 617, 1142

\bibitem[{{Schmeja} \& {Kimeswenger}(2001)}]{schmeja01}
{Schmeja}, S. \& {Kimeswenger}, S. 2001, \aap, 377, L18

\bibitem[{{Sch{\"o}ier}(2007)}]{schoier07}
{Sch{\"o}ier}, F.~L. 2007, in Astronomical Society of the Pacific Conference
  Series, Vol. 378, Why Galaxies Care About AGB Stars: Their Importance as
  Actors and Probes, ed. {F.~Kerschbaum, C.~Charbonnel, \& R.~F.~Wing}, 216

\bibitem[{{Schwarz} {et~al.}(1997){Schwarz}, {Aspin}, {Corradi}, \&
  {Reipurth}}]{schwarz97}
{Schwarz}, H.~E., {Aspin}, C., {Corradi}, R.~L.~M., \& {Reipurth}, B. 1997,
  \aap, 319, 267

\bibitem[{{Smith} {et~al.}(2005){Smith}, {Balick}, \& {Gehrz}}]{smith2-05}
{Smith}, N., {Balick}, B., \& {Gehrz}, R.~D. 2005, \aj, 130, 853

\bibitem[{{Smith} \& {Gehrz}(2005)}]{smith05}
{Smith}, N. \& {Gehrz}, R.~D. 2005, \aj, 129, 969

\bibitem[{{Soker}(2001)}]{soker01}
{Soker}, N. 2001, \apj, 558, 157

\bibitem[{{Solf}(2000)}]{solf00}
{Solf}, J. 2000, \aap, 354, 674

\bibitem[{{Swings} \& {Andrillat}(1979)}]{swings79}
{Swings}, J.~P. \& {Andrillat}, Y. 1979, \aap, 74, 85

\bibitem[{{Torres-Peimbert} {et~al.}(2010){Torres-Peimbert}, {Arrieta}, \&
  {Bautista}}]{torres10}
{Torres-Peimbert}, S., {Arrieta}, A., \& {Bautista}, M. 2010, \rmxaa, 46, 221

\bibitem[{{van Winckel}(2003)}]{winckel03}
{van Winckel}, H. 2003, \araa, 41, 391

\bibitem[{{Zweigle} {et~al.}(1997){Zweigle}, {Neri}, {Bachiller}, {Bujarrabal},
  \& {Grewing}}]{zweigle97}
{Zweigle}, J., {Neri}, R., {Bachiller}, R., {Bujarrabal}, V., \& {Grewing}, M.
  1997, \aap, 324, 624

\end{thebibliography}

\end{document}